\documentclass[preprint,12pt]{elsarticle}
\usepackage[english]{babel}
\usepackage[usenames,dvipsnames]{color}
\usepackage{amssymb}
\usepackage{amsmath}
\usepackage{enumerate}
\usepackage{graphicx}
\usepackage{bm}
\usepackage{hyperref}
\usepackage{makeidx}
\usepackage{textcomp}
\usepackage[usenames,dvipsnames]{color}
\usepackage{comment}

\begin{document}

\title{Correlation between slip precursors and topological length scales at the onset of frictional sliding}

\author[lo]{Gianluca Costagliola}
\ead{gianluca.costagliola@epfl.ch}
\author[to]{Federico Bosia} 
\ead{federico.bosia@polito.it}
\author[tn,qm]{Nicola M. Pugno\corref{cor1}}
\ead{nicola.pugno@unitn.it}

\address[lo]{Civil Engineering Institute, \'Ecole Polytechnique F\'ed\'erale de Lausanne - EPFL, 1015 Lausanne, Switzerland}
\address[to]{ DISAT, Politecnico di Torino, Corso Duca degli Abruzzi 24, 10129 Torino, Italy}
\address[tn]{Laboratory for Bioinspired, Bionic, Nano, Meta Materials \& Mechanics, Department of Civil, Environmental and Mechanical Engineering, University of Trento, Via Mesiano 77, 38123 Trento, Italy}
\address[qm]{School of Engineering and Materials Science, Queen Mary University of London, Mile End Road, London E1 4NS, UK}

\cortext[cor1]{Corresponding author}

\begin{abstract}
Understanding the interplay between concurrent length scales is a fundamental issue in many problems involving friction between sliding interfaces, from tribology to the study of earthquakes and seismic faults. On the one hand, a macroscopic sliding event is preceded by slip precursors with a characteristic propagation length scale. On the other hand, the emergent frictional properties can be modified by surface patterning depending on their geometric length scale. This suggests
that macroscopic sliding of structured surfaces is governed by the interplay between the length scale of the slip precursors and those characterizing the geometric features. In this paper, we investigate these aspects by means of numerical simulations using a two-dimensional spring-block model. We discuss the influence of the geometric features on the occurrence and localization of slip precursors, extending the study to interfaces characterized by two geometric length scales. We find that different types of detachment sequences are triggered by specific surface structures, depending on their scales and relation to sliding direction, leading to a macroscopically smooth transition to sliding in the case of hierarchical and/or anisotropic features. These concepts could be exploited in devices switching from static to dynamic sliding, and can contribute to an improvement in the understanding and interpretation of seismic data.
\end{abstract}

\maketitle

\section{Introduction}

The onset of frictional sliding motion between two dry surfaces displays an intriguing interplay of effects that have been investigated in many recent works. The existence of precursors, i.e. local slip events before the onset of sliding, has been verified experimentally \cite{Rubin_PRL,Maegawa_precurs,Katano} and captured by many numerical models \cite{UrbakhBouch_PRL,Kammer,KammerRadiguet,Norw_pnas,ZapperiTaloni,Roch}. The propagation of wavefronts with different velocities has also been observed both in experiments 
\cite{Rubin_Nature,Rosak_Science,Rosak_SimExp,Fineberg} and in simulations \cite{Braun,Tromborg,Bouchbinder_PRE,Norw1,Norw2,SvetMolinar}, and analogies with fracture mechanics have been highlighted and exploited to quantitatively investigate this aspect of frictional phenomena \cite{Barras,SvetFineberg,Norw3,Molinari_frac}. 

On nominally flat surfaces, the regions where slip precursors occur is influenced by the tangential driving force: in \cite{Rubin_PRL} they were observed at the trailing edge and their length, i.e. the their propagation distance, has been predicted by means of fracture mechanics models \cite{Molinari_frac,Fineberg_frac}. The frequency of these precursors, which redistribute the surface stress in a non-uniform way and induce a change of the contact area, grows until a macroscopic rupture front nucleates and propagates along the whole surface. The emergent static global friction properties, i.e. the first maximum load level reached by the total tangential force in a sliding test, is largely determined by these effects, as recently shown in \cite{GvirFineberg}.

These studies are important not only to increase our understanding of basic friction phenomena taking place at the interface between solids, but also for applications in biology and \cite{LabonteInsects,GorbSnake,Tramsen}, in engineering and materials science \cite{Etsion,Gachot,Rosenkranz}, and in geophysics, to interpret the signals recorded during the preparatory phase preceding an earthquake \cite{Dieterich,Rice93,LapustaRice,Bouchbinder_geo,Lambert}.

This body of literature demonstrates the importance of studying slip precursors and the dynamics underlying stress redistribution before and during the transition to sliding. However, the understanding of these phenomena is often complicated by the presence of multiple length scales at the interfaces. Thus, it is crucial to investigate how the interplay between length scales affects the formation and propagation of slip precursors.

In \cite{Our2D}, we investigated how the macroscopic static friction coefficient is affected by two-dimensional surface structures, which give rise to these non-trivial phenomena. We observed a non monotonic behavior of the static friction coefficient with the surface length scales and non linear effects for anisotropic shapes. Moreover, we showed that the global static friction is not only determined by the non-uniform tangential stress distribution before the sliding phase, but also by the dynamics of the transition, i.e. the propagation of the slip front. An example of this is the case of a patterned surface with straight narrow grooves. This configuration has been analysed numerically and experimentally in the literature, in particular in the case of laser textured surfaces \cite{Prodanov, Gachot_cit, Rosenkranz_cit, Valeri}, demonstrating that patterning reduces friction, but with a non-trivial dependence on system geometric parameters. In our study, considering an elastic patterned surface over a rigid flat substrate, we have found that static friction is greater when the grooves are perpendicular to the sliding direction rather than parallel to it. However, the tangential stress on the edges before sliding is larger in the former case, so that we would expect a smaller static friction. This apparent paradox is explained by the subsequent dynamic evolution. When the motion begins, grooves aligned with the sliding direction favor the rupture front propagation, so that the global tangential force peak is smaller. This example illustrates how the stress redistribution after every local slip event is fundamental to understand how a surface structure modifies the onset of dynamic motion.

In this paper, we further investigate these concepts from a more general point of view. We focus on determining the localization of slip precursors and the propagation of the rupture waves, which determine the emergent static friction properties. Due to the stress concentrations on the edges of surface patterns, during the rupture process there is a time
sequence of localized slip precursors leading to the sliding of the whole contact region. The length scale of the patterns has a non-trivial interplay with the occurrence and location of slip precursors, with different regimes in which the transition from a static to a dynamic phase, from sharp becomes smooth. We show how the interplay between
slip precursors and two geometric length scales affect the macroscopic sliding event. We deal with this problem by means of numerical simulations of the two-dimensional spring-block model introduced in \cite{Our2D}, allowing to take into
account simultaneously longitudinal and tangential stress on the edges of surface features.

The paper is organized as follows: in section \ref{sec_mod} we present the model. In section \ref{sec_nopat},
we consider the case of non-patterned surfaces, to clarify the general dynamic of the model and to obtain benchmark results. In section \ref{sec_pat} we consider square pillar patterns to explain the interplay between the pattern length scale and the slip precursors. In section \ref{sec_hpat}, we introduce the hierarchical patterning and in section \ref{sec_dis} and \ref{sec_con} we provide the discussion and conclusions.

\section{Model} \label{sec_mod}

\begin{figure}[h]
\begin{center}
\includegraphics[scale=0.4]{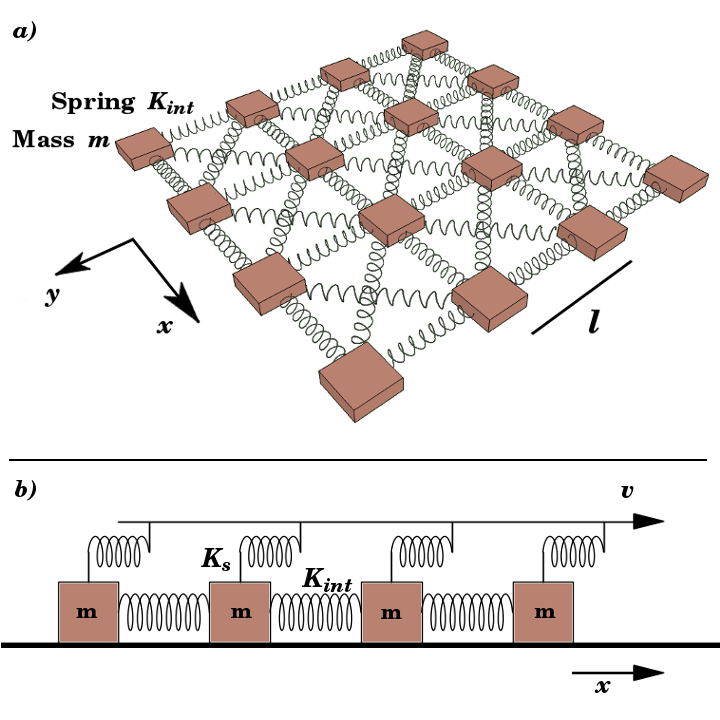}
\caption{\footnotesize a) Discretization of a square surface into a 2-D spring-bock model, showing the mesh of the internal springs $K_{int}$. The shear springs ($K_s$) attached above the blocks are not shown. b) Side view showing the slider moving at constant velocity $v$ and the shear springs. }
\label{model}
\end{center}
\end{figure}

We adopt the formulation of the spring-block model introduced in \cite{Our2D}, considering an elastic 
material sliding over a non-deformable rigid substrate. The elastic surface is discretized into elements of mass $m$, each connected by springs to the eight first neighbors and arranged in a regular square mesh (figure \ref{model}a) with $N_x$ ($N_y$) blocks along the $x$($y$) axis at a distance $l_x$($l_y$). In order to obtain the equivalence of this spring-mass system with a homogeneous elastic material of Young's modulus $E$, the Poisson's ratio is fixed to $\nu=1/3$ \cite{Absi}, 
$l_x=l_y\equiv l$ and $K_{int} = 3/4 E l_z$, where $l_z$ is the thickness of the 2-D layer and $K_{int}$ is the stiffness of the springs connecting the four nearest neighbors of each block, i.e. those aligned with the axes. The stiffness of the diagonal springs must be $K_{int}/2$. 

The internal elastic force on the block $i$ exerted by the neighbor $j$ is 
$\textbf{F}_{int}^{(ij)}  = k_{ij} (r_{ij}-l_{ij}) (\textbf{r}_j-\textbf{r}_i)/r_{ij}  $, where $\textbf{r}_i$, $\textbf{r}_j$ are the position vectors of the two blocks, $r_{ij}$ is the modulus of their distance, $l_{ij}$ is the modulus of the rest distance and $k_{ij}$ is the stiffness of the spring connecting them.

All the blocks are connected by springs of stiffness $K_{s}$ to the slider that is moving at constant velocity $v$ in the $x$ direction (figure \ref{model}b). Hence, the shear force exerted by the slider on the block $i$ is $\textbf{F}_{s}^{(i)} = K_s ( \textbf{v}t +\textbf{r}_i^{0} - \textbf{r}_i )$, where $\textbf{r}_i^{0}$ is the initial rest position of the block and $\textbf{v} = (v,0)$ is the slider velocity vector. The total driving force on $i$ can be defined as $\textbf{F}_{mot}^{(i)} = \sum_{j} \textbf{F}_{int}^{(ij)} + \textbf{F}_{s}^{(i)}$. The stiffness $K_s$ can be expressed as a function of the macroscopic shear modulus $G = 3/8 E$, so that by simple calculations we obtain $K_s = K_{int} l^2/l_z^2$. In the following, for simplicity we fix $l_z=l$. Thus, the elementary mass is $m=\rho l^3$, where $\rho$ is the material density. Moreover, we consider only a square mesh, i.e. $N_x = N_y \equiv N$.

In the spring-block model, a viscous damping force, proportional to the velocity, is added to eliminate artificial block oscillations, i.e. $\textbf{F}_{d}^{(i)} = - \gamma m \dot{\textbf{r}_i}$, where $\gamma=5 \; ms^{-1}$ is the damping frequency.

The interaction with the substrate is modeled as described in \cite{Our2D,Our1D}, i.e. as an Amontons-Coulomb friction force with a statistical distribution on the friction coefficients: while the block $i$ is at rest, the friction force $\textbf{F}_{fr}^{(i)}$ opposes the total driving force, i.e. $\textbf{F}_{fr}^{(i)} = - \textbf{F}_{mot}^{(i)}$, up to a threshold value $F_{fr}^{(i)} = {\mu_s}_i \; F_n^{(i)}$, where ${\mu_s}_i$ is the static friction coefficient and $F_n^{(i)}$ is the normal force on $i$. When this limit is exceeded, a constant dynamic friction force opposes the motion, i.e. 
$\textbf{F}_{fr}^{(i)} = - {\mu_d}_i \; F_n^{(i)} \widehat{\dot{\textbf{r}_i}}$, where ${\mu_d}_i$ is the dynamic friction coefficient and $\widehat{\dot{\textbf{r}_i}}$ is the velocity direction of the block. Given the applied pressure $P$, the normal force $F_n^{(i)}$ can be calculated as $F_n^{(i)} = P l^{2}$ and the total normal force is $F_n = P l^{2} N^{2}$.

In the following, we will drop the subscript $s$,$d$ every time the considerations apply to both the coefficients. The microscopic friction coefficients are extracted from a Gaussian statistical distribution to account for the randomness of the surface asperities, i.e. $p({\mu}_i) = (\sqrt{2\pi}\sigma)^{-1} \exp{[-({\mu}_i-(\mu)_m)^2/(2\sigma^2)]} $, where $(\mu)_m$ denotes the mean of the microscopic friction coefficients and $\sigma$ is its standard deviation. The mean microscopic static and dynamic friction coefficients are fixed conventionally to $(\mu_s)_m = 1.0(1)$ and $(\mu_d)_m = 0.60(5)$, respectively, where the numbers in brackets denote the standard deviations of their Gaussian distributions.

Thus, the Newton's law for the block $i$ is: $m \ddot{\textbf{r}_i} = \sum_j \textbf{F}_{int}^{(ij)} + \textbf{F}_{s}^{(i)} + \textbf{F}_{fr}^{(i)} + \textbf{F}_{d}^{(i)}$. The overall system of differential equations is solved using a fourth-order Runge-Kutta algorithm. In order to calculate the average of any observable, the simulation must be repeated various times, extracting each time new random friction coefficients. In repeated tests, an integration time step $h=10^{-8} s$ proves to be sufficient to reduce integration errors under the statistical uncertainty in the range adopted for the parameters of the system.

The global emergent friction force can be calculated from the sum of the longitudinal forces exerted by the slider, i.e. $F_{fr}= | \sum_i \textbf{F}_{s}^{(i)}|$, which corresponds to the longitudinal friction force that would be measured in an experiment of linear sliding.

Realistic values are chosen for the physical system parameters, i.e. a Young's modulus $E = 10$ MPa, 
which is typical for a soft polymer or rubber-like material, a density $\rho = 1.2$ g/cm$^{3}$, and a normal pressure $P=0.05$ MPa. We have chosen these values as an average of those used in previous experimental work on these systems \cite{Balestra,Berardo}, but we note that, in the spring-block model, the relevant parameters is the ratio between Young’s modulus and applied pressure, so that our results are qualitatively the same for all larger Young’s moduli, provided that the pressure is also increased. The slider velocity is $v = 0.05 $ cm/s, similar to those adopted in recent experiments with surface textures \cite{Berardo,Scheibert,Maegawa_patt}. The distance between blocks $l$ in the model is an arbitrary parameter representing the smallest surface feature that can be modeled and is chosen by default as $l = 10^{-3}$ cm. 

For discussion on the effect of these approximations, the comparison with other formulations of the literature, and the choices of parameters, we refer the reader to \cite{Our2D}. We remark here that the formulation is appropriate to describe the transition to sliding for slow velocities, assuming a constant real contact area and neglecting any long-term dynamical effects. For this reason, it is more effective in estimating the values of the static friction coefficient with respect to the dynamic ones \cite{Berardo}. Moreover, the current setup cannot capture phenomena related to the vertical stress distributions \cite{Guarino} or wave propagation in the bulk, and those related to the modification of the real contact area \cite{Balestra,Scheibert}. Nevertheless, the model is able to provide a good qualitative understanding at the onset of sliding at relatively small computational costs with respect to other full three-dimensional methods.

In order to simulate the presence of patterning on the surface, we set to zero the friction coefficients of the blocks located in correspondence with detached regions from the rigid plane. These blocks are not considered in contact with the ground, and the total normal force is distributed only on contact blocks. This is a good approximation to find the in-plane stress distribution at least for shallow structures, i.e. when their length along the vertical axis is larger than surface roughness, but sufficiently small to avoid significant effects due to the vertical shear stress, like deformation, tilting or bending of these structures. In \cite{Berardo}, in which surface structures fulfil this condition, this model is able to reproduce the qualitative trends of the static friction coefficients by varying the structure length.

%\clearpage

\section{Results}

\subsection{Non-patterned surface} \label{sec_nopat}

As discussed in \cite{Our2D}, our formulation of the spring-block model (for a non-patterned surface) provides a qualitative description of the onset of the sliding motion. In the following, we will refer to detachment of a single block to indicate its transition from static to dynamic friction.

\begin{figure}[h!]
\begin{center}
\includegraphics[scale=0.55]{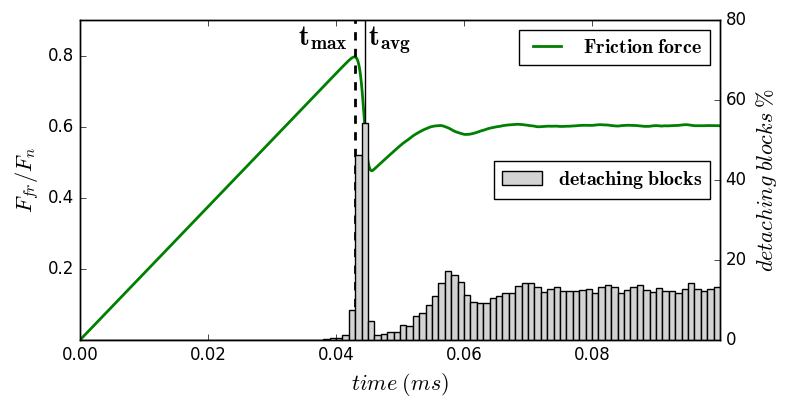}
\caption{\footnotesize Time evolution for the total friction force and percentage of detaching blocks, i.e. blocks switching
from static to dynamic friction in small time ranges, for a non-patterned surface. The vertical dotted lines mark the maximum of the total friction force, occurring at $t_{max}$ and the thin solid line the average detachment time $t_{avg}$. Slip precursors appear before $t_{max}$, while the global detachment event occurs shortly after it. This simulation is performed with default parameter values and $N=120$.}
\label{fig1}
\end{center}
\end{figure}

The typical time evolution of the total friction force is shown in figure \ref{fig1}. Slip precursors occur as single detachment events before the macroscopic sliding phase, corresponding to the weakest microscopic static friction thresholds. From these events,macroscopic rupture waves may nucleate. Parameters like the slider velocity, pressure, or statistical distributions of the local friction coefficients all affect the number of wave fronts, but in any case the time instant 
$t_{max}$ at which the maximum tangential force occurs and the global static coefficient is calculated, corresponds to the onset of  propagation. Once the detachment fronts have traveled across the whole contact surface, a non-uniform distribution of residual strains appears, at the time instant corresponding to the relative force minimum after the first peak in figure \ref{fig1}. These residual strains are removed at the beginning of the dynamic phase \cite{KammerRadiguet}, when localized stick-slip motions of small fractions of the contact area take place.

\begin{figure}[h!]
\begin{center}
\includegraphics[scale=0.55]{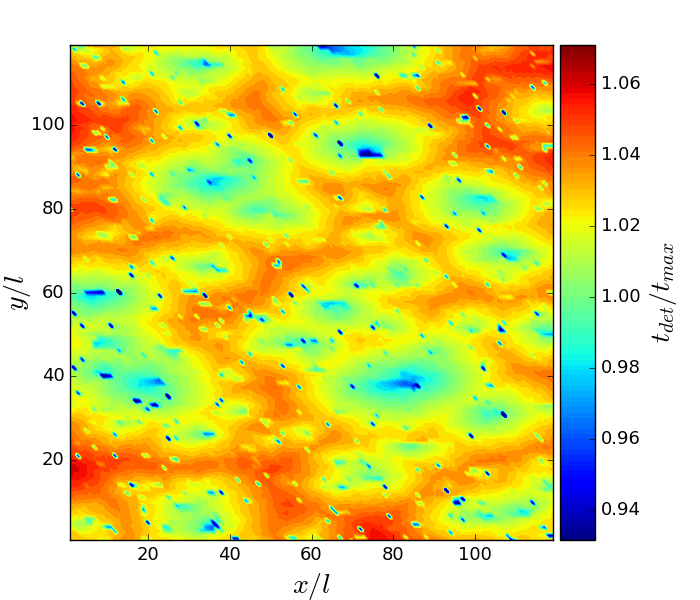}
\caption{\footnotesize Color map of the times $t_{det}$ of the first detachment of the blocks on the surface, corresponding to the time evolution of figure \ref{fig1}. Times are normalized with respect to $t_{max}$, so that blue corresponds to detachment before the force peak and the red to those after that. The slip precursors appear as dark blue spots, not all of which give rise to a detachment nucleation point, e.g. those located in red regions.}
\label{fig2}
\end{center}
\end{figure}

Figure \ref{fig1} displays the corresponding number of blocks detaching over time: the precursors are those occurring before $t_{max}$, while this time instant there is the main sliding avalanche, highlighted by the peak of the number of detachments. The two peaks are clearly separated: $t_{max}$ occurs before the average detachment time calculated over all blocks, namely $t_{avg}$, which corresponds roughly to the propagation instant of the wave fronts over the whole surface. The time sequence of detachments on the two-dimensional surface is graphically illustrated in figure \ref{fig2}. From this we observe that not all the precursors trigger a rupture wave front.

\clearpage

\subsection{Single-scale patterned surfaces} \label{sec_pat}

\begin{figure}[h!]
\begin{center}
\includegraphics[scale=0.25]{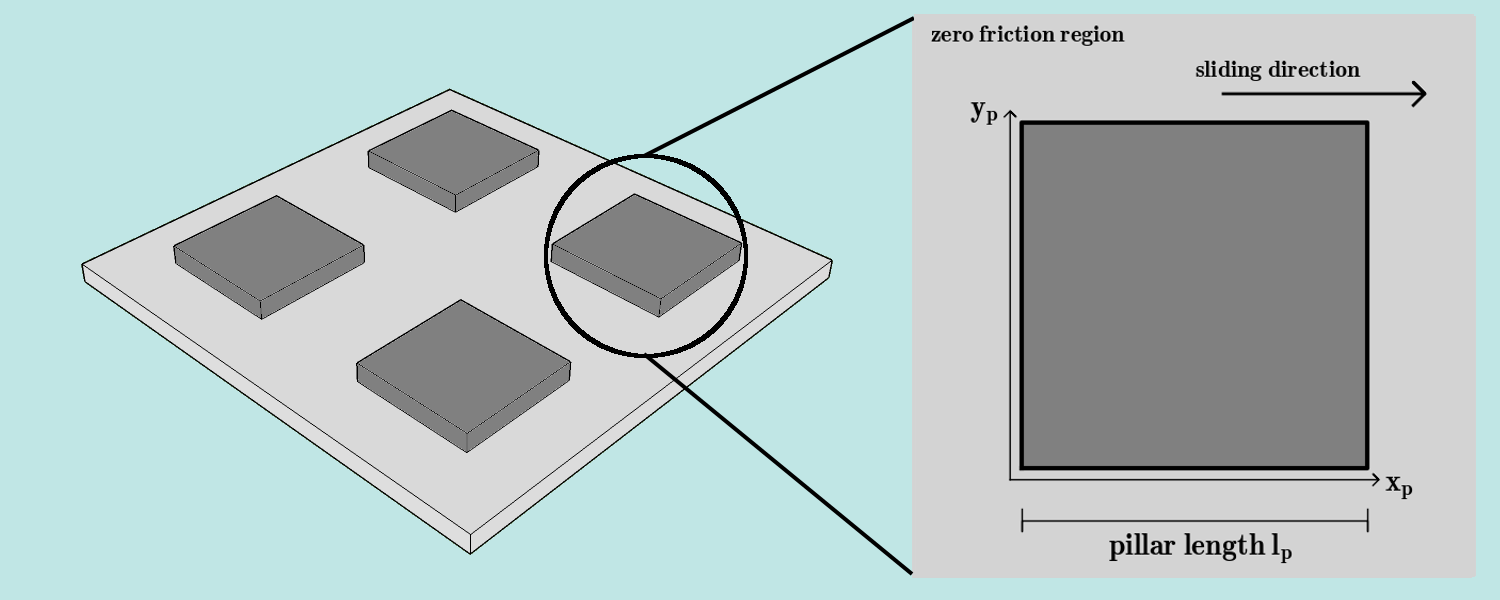}
\caption{\footnotesize Surface patterning using square pillars. A single pillar is highlighted in the enlargement with the coordinates of its points ($x_p,y_p$) adopted in the next plots.}
\label{fig3}
\end{center}
\end{figure}

On non-patterned surfaces, the localization of slip precursors and nucleation points are determined by the statistical distributions of the local friction coefficients. Thus, on average, each point of the surface has the same probability of being a nucleation point, and in repeated simulations the spatial distribution on the surface of the initial detachment time is uniform. This situation changes in the presence of surface patterning. Here, we consider square pillars, which are a simple configuration to investigate these effects from a qualitative point of view.

In order to simulate these structures, we set to zero the local friction coefficients of the blocks not corresponding to the pillar regions. We indicates with $l_p$ and $l_s$ the pillar size and spacing, respectively, which can be normalized by the elementary block distance $l$, so that we can define $n_p \equiv l_p/l$ and $n_s \equiv l_s/l$ corresponding to the pillar and spacing length expressed in terms of number of blocks. Figure \ref{fig3} shows an example of a surface portion characterized by equally spaced square pillars. 

The distance between surface edges and the pillars close to them is $l_s/2$ to reduce boundary effects. We have verified in preliminary tests that, with this choice, all the pillars have the same average statistical properties and are subject to the same stress distributions for $n_p \geq 4$. The case $n_p \leq 3$ is peculiar, since the number of blocks on the pillar is too small to induce a non-trivial stress distribution. In this case, the transition to sliding is similar to that for a non-patterned surface, so that in the following analysis we restrict to the cases $n_p \geq 4$. 

\begin{figure}[h!]
\begin{center}
\includegraphics[scale=0.6]{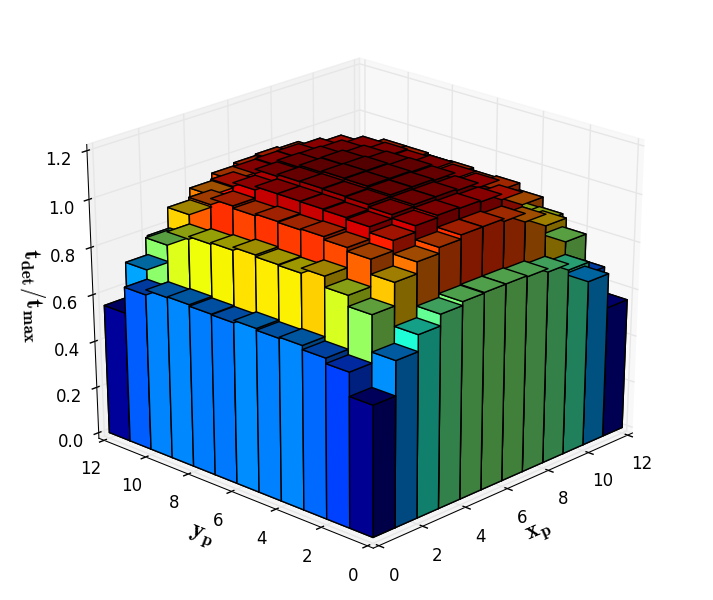}
\caption{\footnotesize Average first-detachment time $t_{det}$ (normalized with respect to $t_{max}$) of the blocks on a single pillar of size $n_p=20$.}
\label{fig4}
\end{center}
\end{figure}

The non-uniform surface stress distribution occurring on the surface of a single pillar before sliding induces a specific detachment time sequence, as shown in figure \ref{fig4}. First, regions on the corners detach, then those on the boundaries perpendicular to the sliding direction, followed by those on the parallel boundaries. This is because the longitudinal stress exerted on orthogonal edges is larger than in-plane shear stress along parallel ones. This sequence is repeated for the blocks adjacent to those on the borders. These detachments occur before the maximum of the friction force is reached and the global sliding begins, so that they can be considered sliding precursors. Finally, detachment waves are triggered from nucleation points located near the edge and propagate towards the central part of the pillar, which is the last part detaching. The number of wave fronts crossing the pillar and the exact location of the nucleation points is influenced by the random distribution of the local friction coefficients, but the described detachment sequence always occurs, 
independently of statistical properties.

The sequence reflects the initial stress distribution and its redistribution after the first detachments, but the macroscopic consequence is the inversion between $t_{max}$ and $t_{avg}$, i.e. the force peak always occurs after the average detachment time. This implies that the sliding precursors are not only a few isolated blocks, but a whole region localized around the edge of the pillar. This is a qualitative change of the dynamics occurring at the onset of sliding motion, which is reflected in the shape of the force time evolution curve that is no longer a sharp peak, and becomes rounded. 

\begin{figure}[h!]
\begin{center}
\includegraphics[scale=0.6]{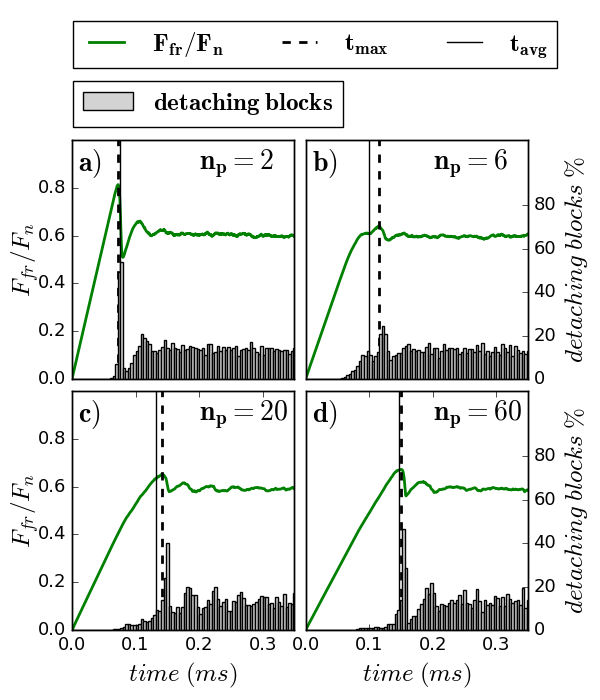}
\caption{\footnotesize Time evolution of the friction force and block detachments for various $n_p$ values. In the case $n_p=2$, the dynamics of the transition to sliding is similar to that of non-patterned surfaces, while in the other cases the detachment sequence described in section \ref{sec_pat} occurs.}
\label{fig5}
\end{center}
\end{figure}

This can be observed in figure \ref{fig5}, which illustrated the time evolution of the force and the number of detachments for various values of  $n_p$. While for $n_p=2$ the behavior is similar to a non-patterned surface, for $n_p \geq 6$ the sliding precursors are numerically relevant and occur for a larger time interval before $t_{max}$. For very large pillars, e.g. $n_p=60$, the static friction coefficient increases again and the transition is sharper. This is because the region 
characterized by sliding precursors, localized around the edges of the pillars, becomes negligible with respect to the surface area of the central part, which is involved in the main detachment front after the time $t_{max}$. 

The characteristic length of the slip precursors on the pillar sides in orthogonal directions is not symmetric due to the preferential sliding direction. This asymmetric behavior have been observed in recent experiments \cite{Scheibert}. In the 2-D spring-block model, this causes different detachment processes at orthogonal sides of the pillar. 

If $\tau_{1}$,$\tau_{2}$ are the detachment times of two points $x_1$ and $x_2$ located on the same pillar side, we can calculate the time correlation function $C(x_1,x_2)\equiv \langle \tau_{1} \tau_{2}\rangle- \langle \tau_{1} \rangle 
\langle \tau_{2} \rangle$, where $\langle...\rangle$ denotes the average over repeated simulations. If the correlation is zero, the detachment times in repeated simulations are uncorrelated. This implies that, despite $\tau_{1}$,$\tau_{2}$ have the same average value, they detach independently of each other. Instead, in the presence of correlation, their detachment could be triggered by the same precursor events involving more blocks. The precursor length can be estimated as the 
distance over which a significant correlation exists.

\begin{figure}[h!]
\begin{center}
\includegraphics[scale=0.7]{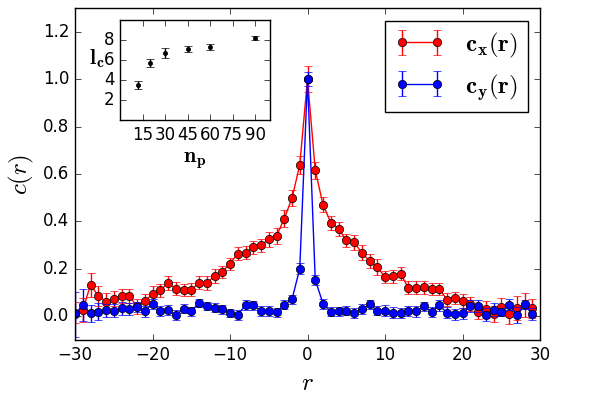}
\caption{\footnotesize Time correlation functions $c_{x}(r)$ and $c_{y}(r)$ on orthogonal sides of the pillar for the case 
$n_p=60$. A point located on a pillar is identified by its coordinates as defined in figure \ref{fig3} $x=(x_p,y_p)$. On the 
side aligned with the sliding direction $x$, the points are $x_1=(n_p/2,0)$ and $x_2=(n_p/2+r,0)$, and $c_{x}(r)\equiv 
C(x_1,x_2)/C(x_1,x_1)$. On the side along $y$, the points are $y_1=(0,n_p/2)$ and $y_2=(0,n_p/2+r)$, and $c_{y}(r)\equiv 
C(y_1,y_2)/C(y_1,y_1)$. In the inset, the estimates of the characteristic correlation length $l_c$ as a function of the pillar size $n_p$ obtained through the fit of $c_{x}(r) \sim e^{-r/l_c}$.}
\label{fig6}
\end{center}
\end{figure}

In figure \ref{fig6}, we report the correlation function calculated from the central point of each side as a function of the distance along the same side. In symbols, on sides along $x$, the time correlation function are calculated between the points $x_1=(n_p/2,0)$ and $x_2=(np/2+r,0)$ as $c_{x}(r)\equiv C(x_1,x_2)/C(x_1,x_1)$, which is the normalized correlation function. The same definition applies for the side along $y$ between the points $y_1=(0,n_p/2)$ and $y_2=(0,np/2+r)$, with
$c_{y}(r)\equiv C(y_1,y_2)/C(y_1,y_1)$. These correlation functions are shown in figure \ref{fig6} for various $r$ values: while $c_{y}(r)$ decreases rapidly, $c_{x}(r)$ displays a correlation over a wide range of values. From this, we deduce that on sides aligned with the sliding direction slip precursors have a non-negligible characteristic correlation length. Although edges parallel to the sliding direction detach sligthly after the perpendicular ones, their slip involves a larger number of blocks.

The correlation functions decreases exponentially, i.e. $c_{x}(r) \sim e^{-r/l_{c}}$, where $l_c$ is the characteristic correlation length, that can be estimated by fitting the correlation functions. In the inset of figure \ref{fig6}, we report the fit results of $l_c$ for various pillar sizes $n_p$, obtained by averaging the correlation functions. It is difficult to estimate $l_c$ for $n_p \lesssim 10$ since there are few blocks to fit, however we can deduce that $l_c/l 
\gtrsim 3$ and that it is slightly increasing with $n_p$.  

This means that when the characteristic correlation length is comparable to the total length of the pillars, a slip precursor can lead to the detachment of the whole side of the pillar, leading to a faster detachment process and a globally smaller static friction. This explains why the smallest static friction is obtained for an specific intermediate value of $n_p$, as can be observed in figure \ref{fig5}, and the force peak at the transition is considerably more rounded.

Thus, this formulation of the spring-block model predicts an optimal length scale of the pillars which maximizes the occurrence of sliding precursors, so that the static friction is minimized and the transition to sliding is facilitated. One could argue that this features is inherent to the discretization length of the model. However, such minimum value for intermediate values of the pattern sizes is observed in experimental results \cite{Varenberg,Li,Berardo}. In real systems, there is a characteristic length scale of the precursors \cite{Molinari_frac}\cite{Fineberg_frac} which must be taken into account and compared to the length scale of the patterns. When the precursor length scale is greater or much smaller than the pattern size, the transition is qualitatively similar to that of a non-patterned surfaces, although the effective emergent value of the static friction coefficient can be different due to the different stress distributions. In the intermediate regime, i.e. for precursor length scales comparable to the pattern size, the slip precursors involve a 
relevant fraction of the contact surface and the detachments are distributed over a longer time interval around the maximum of the friction force. This leads to a longer duration of the transition between static and dynamic friction.

\clearpage

\subsection{Multiscale patterned surfaces} \label{sec_hpat}

\begin{figure}[h!]
\begin{center}
\includegraphics[scale=0.25]{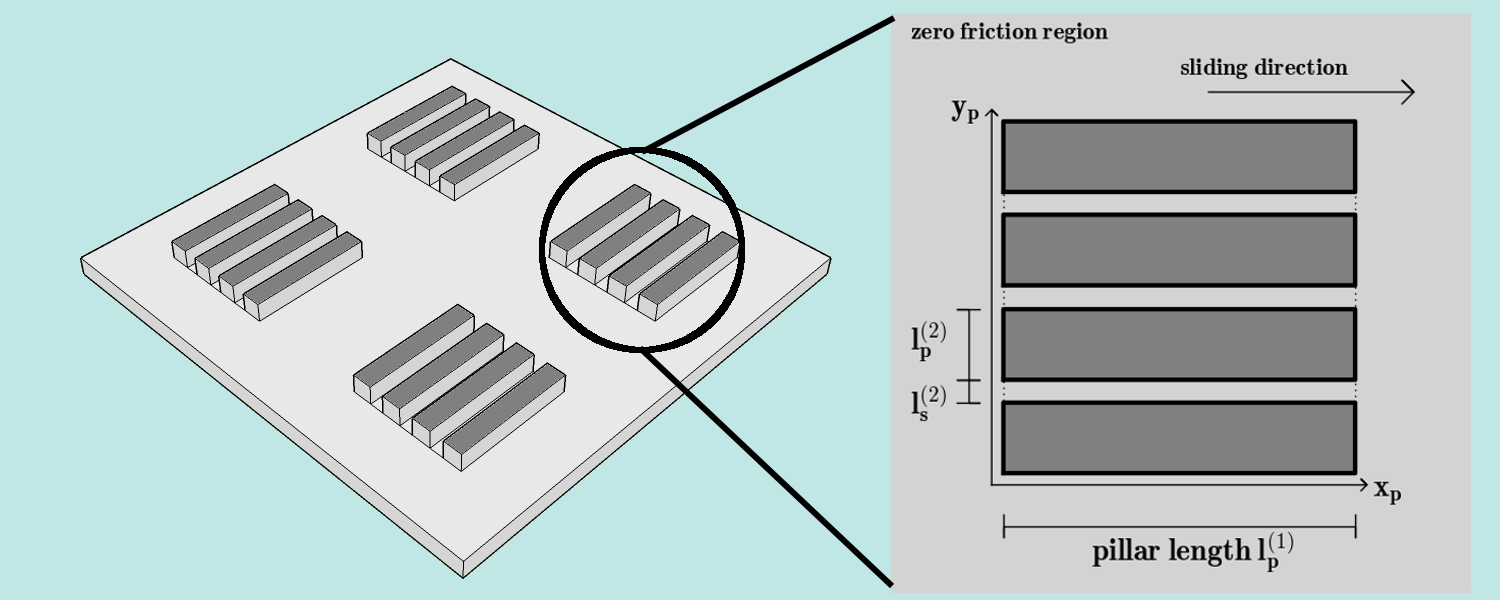}
\caption{\footnotesize Second level hierarchical patterning with high aspect ratio structures. In the enlargement, a single 
first level single pillar and second level structures in the adopted coordinate system.}
\label{fig8}
\end{center}
\end{figure}

Next, we consider the case in which the patterning geometry (i.e. the square pillars discussed in section \ref{sec_pat}) are split into smaller sub-structures, giving rise to the presence of two different length scales in the patterning structure. The simplest structure to consider is a pillar characterized by high-aspect ratio longitudinal or transversal 
grooves, as shown in figure \ref{fig8}. We simulate these configurations again by setting to zero the 
friction coefficients of blocks corresponding to the grooves. We adopt the same notation of section \ref{sec_pat}, i.e. a point on a pillar is denoted with its coordinates $(x_p,y_p)$. Length and spacing of the first level pillar are denoted with 
$l_p^{(1)}$ and $l_s^{(1)}$, respectively. The structure sizes of the second-level, in this case pawls and grooves, are denoted with $l_p^{(2)}$ and $l_s^{(2)}$, respectively. Any length $l_{i}$ with a generic subscript $i$ can be normalized by the elementary length $n_{i} \equiv l_{i}/l$.

This type of multiscale (hierarchical) configuration combines the effects analyzed in the previous section due to the precursors appearing along the sliding direction, with another characteristic length due to the second level structures. As observed in \cite{Our1D,Lucas}, a hierarchical configuration allows to distribute mechanical stresses more uniformly over the surface and, in this case, increases the total effective length of the edges, i.e. the regions giving rise to slip precursors. Thus, there is a precise rupture time sequence as in the single-level patterns, starting from blocks located on the transversal side of the pillar, followed by those on the longitudinal external one. Shortly after these, there is the detachment of the blocks located along the longitudinal internal sides, i.e. along the second level grooves. The last stage of detachment takes place in the internal regions of the pawls.

\begin{figure}[h!]
\begin{center}
\includegraphics[scale=0.6]{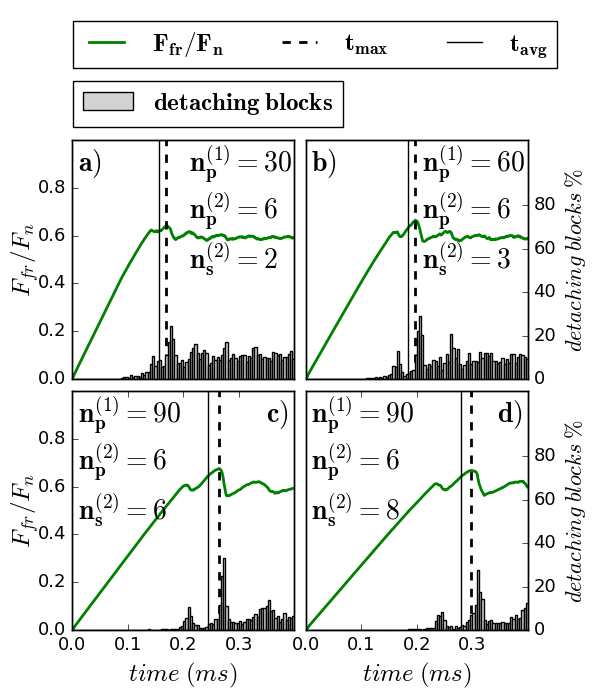}
\caption{\footnotesize Time evolution of the friction force and block detachment for various hierarchical configurations. 
the transition from static to dynamic friction is characterized by a double peak, corresponding to the distinct 
detachments of boundary and central regions of the structures.}
\label{fig9}
\end{center}
\end{figure}

In the time evolution of the longitudinal friction force, the transition from static to dynamic friction is characterized by two load peaks separated by an intermediate quasi-static phase (see figure \ref{fig9}). The geometry of these configurations divides the contact surface into two parts of equal size with distinct dynamics: the edge regions where slip precursors originate, and the internal core of the contact structures. The external edges of the pillars are subjected to larger 
longitudinal stresses than the internal ones along the second-level grooves. For this reason, the former detach before the latter, but the correlation functions along the longitudinal edges display the same qualitative shape shown in section \ref{sec_pat}. The typical correlation length is also the same. Thus, all the edges along the second-level grooves are characterized by slip precursor with non-negligible length. This lead to a faster detachment of the boundary regions and, consequently, to the first load peak.

The time interval between the first and the second peak is controlled by the difference between the length scales. The global static friction, which is determined by the force reached at the second load peak, can be manipulated by modifying the pillar length $l_p^{(1)}$, although in general all the factors analyzed in \cite{Our2D} must be considered.

The two load peaks appear only if the second level grooves are aligned with the sliding directions. On the contrary, if the sliding direction is along the $y$-axis, most of the precursors are uncorrelated so that there is no large detachment determining the first load peak observed in figure \ref{fig9}. Thus, this effect is due to two fundamental factors: the concurrent presence of multiple characteristic length scales, $l_s^{(2)}<<l_s^{(1)}$ (hierarchical structure), and the anisotropic shape of the second-level grooves exploiting the occurrence of slip precursors with non-negligible length along the longitudinal edges.

The results obtained for a simple two-level structure can be generalized to multiple length scales and more complex geometries or to other surface patterns, provided that a hierarchical organization is present. Hierarchy is necessary but not sufficient, since the shape of the surface patterns and the alignment with the sliding direction must be designed to maximize the occurrence of precursor length. An anisotropic shape, similar to that of the second-level patterns considered here, is another fundamental factor to obtain such non-trivial frictional behaviors. We expect that these effects remain present when adding further hierarchical levels, although other physical mechanisms should be considered beyond the current model at smaller length scales.

\section{Discussion and relevance to seismic phenomena}\label{sec_dis}

The described non-trivial behavior described in the previous section could be exploited in applications involving patterned surfaces for friction control, e.g., whenever a rapid transition from static to dynamic friction is required while maintaining a smooth force behavior. In the phase between load peaks such as those appearing in Fig. 9, the system is in a static friction phase, since the internal regions of the patterning are still at rest, while the regions involved by precursors events undergo a stick-slip motion after the first detachment. However, the whole contact surface is in a stressed state, involving a combination of inhomogeneous stress and partial detachment, which can lead to full detachment in the presence of a small additional force. These effects could also be relevant for understanding the tribology of structured surfaces, including in biological systems. For example, in recent years, the pre-stressed state of the gecko paw has been investigated and identified as one of the key factors responsible for its capability to detach easily despite its strong static adhesion \cite{Gecko_Gao}. 

Additionally, the results discussed in this paper are consistent with a large body of seismic data and could provide elements for its better understanding and interpretation. Starting from the seminal paper by Dieterich \cite{Dieterich}, earthquake precursors have been linked to preseismic fault creep and correlated to geometric dimensions of earthquake source zones. Preseismic slip has been found to be an intrinsic part of the process of homogenization of stress along a fault leading to unstable (seismic) slip \cite{Papadimitri}. Precursor activity has been identified as one of the key ingredients in the field of earthquake prediction \cite{Sykes}, with much discussion on whether the probability for large earthquake occurrence is highest during time periods of smaller event activation or quiescence \cite{Rundle}. 

Analysis of seismic data sets has highlighted the presence of a complex nucleation phase, characterized by slip events distributed over long times \cite{Linde_Aseismic, Schmittbuhl_Izmit, Schmittbuhl_Precurs, Hasegawa} and sequences of localized precursor events have been observed \cite{Ellsworth_1, Ellsworth_2, Ellsworth_3}. Similar conclusions have also been obtained from rigorous earthquake models involving sliding faults governed by rate and state friction laws \cite{LapustaRice,Lapusta}. The relevance of slip precursors to characterize and possibly to predict macroscopic events has also been highlighted by discrete element models \cite{Ferdowsi} and by spring-block models \cite{Brown_SB2D,Mori_1,Mori_2}. 

Our analysis introduces the possibility of correlating the observed phenomenology of seismic events to the geometrical features of the corresponding fault regions, possibly characterising them statistically. In particular, according to our analysis, the presence of an extended precursor phase could be correlated to multiscale fault geometry, involving characteristic length scales of which the fundamental one is expected to be of the order of the precursor correlation lengths. Our analysis on the effects of multiscale hierarchical geometrical features are consistent with studies on the implications on earthquake nucleation of geometric irregularity of the rupturing surfaces in fault lines \cite{Ohnaka}, including when the geometry involves multiscale heterogeneity \cite{Aochi}. Our results also indicate that an extended precursor phase is linked to the presence of pre-existing faults with inhomogeneous stress distributions, reflecting the fact that stress transfer and stress triggering phenomena have been invoked in the past for the onset of seismic activity \cite{Stein}. Specific aspects related to the application of our model to seismic problems will be the focus of future work.

\clearpage

\section{Conclusions} \label{sec_con}

Using a 2-D spring-block model we have investigated the transition from static to dynamic friction in the presence of hierarchical multiscale surface structures. The occurrence of slip precursors and how they affect the global emergent behavior have been discussed. In the case of nominally flat surfaces, isolated precursor events occur before global detachment, leading to the nucleation of a macroscopic rupture front, whose propagation over the whole surface determines the global static friction force.

In the presence of surface patterns, non-uniform surface stress distributions before sliding lead to different detachment sequences , and the macroscopic rupture wave is preceded by multiple precursor events originating from the edges orthogonal and parallel to the sliding direction. A characteristic length emerges, which can be estimated from the time correlation function, affecting the global transition to sliding: when the patterning size and the length of the precursors are of the same order, static friction is smaller and the transition load peak becomes smoothed. 

In the presence of multiple scale levels in the patterning, the effect is replicated at all size scales, leading to a detachment phase characterized by multiple distinct peaks between which the surface is in a static friction condition but also in a stressed state, allowing a rapid but smooth transition to sliding. We show that this complex behavior emerges in the presence of both multiscale and appropriate anisotropic features in the surface patterning, and present a specific example of surface texture with these characteristics. Various other types of surface patterns could be designed to tune the specific behavior in the transition to sliding. These could be exploited to optimize the frictional properties of artificial devices switching continuously from static to dynamic sliding. Moreover, the discussed mechanisms could be responsible for the observed phenomenology in seismic precursor events and could be helpful in predicting earthquake events, based on the deduced average stress state of the involved fault lines.

\section*{Acknowledgments}
FB and NMP are supported by the European Commission under the FET Open ``Boheme'' grant no. 863179.  GC has received funding from the European union's Horizon 2020 research and innovation programme under the Marie Sklodowska-Curie grant agreement no. 754462. The authors acknowledge the Province of Trento for the financial support provided with the LP 6/99 for the "Meta-F" project.

Computational resources were provided by hpc@polito (http://www.hpc.polito.it) and by the Centro di  Competenza  sul Calcolo  Scientifico  (C3S)  of  the  University of  Torino  (c3s.unito.it)

\clearpage

\bibliographystyle{unsrt}
\bibliography{references}

\begin{thebibliography}{10}

\bibitem{Rubin_PRL}
S.M. Rubinstein, G.~Cohen, and J.~Fineberg.
\newblock Dynamics of precursors to frictional sliding.
\newblock {\em Phys. Rev. Lett.}, 98:226103, 2007.

\bibitem{Maegawa_precurs}
S.~Maegawa, A.~Suzuki, and K.~Nakano.
\newblock Precursors of global slip in a longitudinal line contact under
  non-uniform normal loading.
\newblock {\em Tribol. Lett.}, 38:3, 2010.

\bibitem{Katano}
Y.~Katano, K.~Nakano, K.~Otsuki, and H.~Matsukawa.
\newblock Novel friction law for the static friction force based on local
  precursor slipping.
\newblock {\em Sci. Rep.}, 4:6324, 2014.

\bibitem{UrbakhBouch_PRL}
E.~Bouchbinder, E.A. Brener, I.~Barel, and M.~Urbakh.
\newblock Slow cracklike dynamics at the onset of frictional sliding.
\newblock {\em Phys. Rev. Lett.}, 107:235501, 2011.

\bibitem{Kammer}
D.S. Kammer, V.A. Yastrebov, P.~Spijker, and J.F. Molinari.
\newblock On the propagation of slip fronts at frictional interfaces.
\newblock {\em Tribol. Lett.}, 48:27, 2012.

\bibitem{KammerRadiguet}
M.~Radiguet, D.S. Kammer, P.~Gillet, and J.F. Molinari.
\newblock Survival of heterogeneous stress distributions created by precursory
  slip at frictional interfaces.
\newblock {\em Phys. Rev. Lett.}, 111:164302, 2013.

\bibitem{Norw_pnas}
J.~Tr\o{}mborg, H.A. Sveinsson, J.~Scheibert, K.~Th\o{}gersen,
  A.~Malthe-S\o{}renssen, and D.S. Amundsen.
\newblock Slow slip and the transition from fast to slow fronts in the rupture
  of frictional interfaces.
\newblock {\em Proc. Natl. Acad. Sci.}, 111:8764, 2014.

\bibitem{ZapperiTaloni}
A.~Taloni, A.~Benassi, S.~Sandfeld, and Zapperi S.
\newblock Scalar model for frictional precursors dynamics.
\newblock {\em Sci. Rep.}, 5:8086, 2015.

\bibitem{Roch}
T.~Roch, E.A. Brener, J.-F. Molinari, and E.~Bouchbinder.
\newblock Velocity-driven frictional sliding: Coarsening and steady-state
  pulses.
\newblock {\em J. Mech. Phys. Solids}, 158:104607, 2022.

\bibitem{Rubin_Nature}
S.M. Rubinstein, G.~Cohen, and J.~Fineberg.
\newblock Detachment fronts and the onset of dynamic friction.
\newblock {\em Nature}, 430:1005, 2004.

\bibitem{Rosak_Science}
K.~Xia, A.J. Rosakis, and H.~Kanamori.
\newblock Laboratory earthquakes: the sub-rayleigh-to-supershear rupture
  transition.
\newblock {\em Science}, 303:1859, 2004.

\bibitem{Rosak_SimExp}
D.~Coker, G.~Lykotrafitis, A.~Needleman, and A.J. Rosakis.
\newblock Frictional sliding modes along an interface between identical elastic
  plates subject to shear impact loading.
\newblock {\em J. Mech. Phys. Solids}, 53:884, 2005.

\bibitem{Fineberg}
O.~Ben-David, G.~Cohen, and J.~Fineberg.
\newblock The dynamics of the onset of frictional slip.
\newblock {\em Science}, 330:211, 2010.

\bibitem{Braun}
O.M. Braun, I.~Barel, and M.~Urbakh.
\newblock Dynamics of transition from static to kinetic friction.
\newblock {\em Phys. Rev. Lett.}, 103:194301, 2009.

\bibitem{Tromborg}
J.~Tr\o{}mborg, J.~Scheibert, D.S. Amundsen, K.~Th\o{}gersen, and
  A.~Malthe-S\o{}renssen.
\newblock Transition from static to kinetic friction: Insights from a 2d model.
\newblock {\em Phys. Rev. Lett.}, 107:074301, 2011.

\bibitem{Bouchbinder_PRE}
Y.~Bar-Sinai, R.~Spatschek, E.~Brener, and E.~Bouchbinder.
\newblock Instabilities at frictional interfaces: Creep patches, nucleation,
  and rupture fronts.
\newblock {\em Phys. Rev. E.}, 88:060403, 2013.

\bibitem{Norw1}
D.~S. Amundsen, J.~Tr\o{}mborg, K.~Th\o{}gersen, E.~Katzav,
  A.~Malthe-S\o{}renssen, and J.~Scheibert.
\newblock Steady-state propagation speed of rupture fronts along
  one-dimensional frictional interfaces.
\newblock {\em Phys. Rev. E}, 92:032406, 2015.

\bibitem{Norw2}
J.~Tr\o{}mborg, H.A. Sveinsson, K.~Th\o{}gersen, J.~Scheibert, and
  A.~Malthe-S\o{}renssen.
\newblock Speed of fast and slow rupture fronts along frictional interfaces.
\newblock {\em Phys. Rev. E}, 92:012408, 2015.

\bibitem{SvetMolinar}
I.~Svetlizky, D.~Pino~Munoz, M.~Radiguet, D.S. Kammer, J.F. Molinari, and
  Fineberg J.
\newblock Properties of the shear stress peak radiated ahead of rapidly
  accelerating rupture fronts that mediate frictional slip.
\newblock {\em Proc. Natl. Acad. Sci.}, 113:542--547, 2016.

\bibitem{Barras}
F.~Barras, M.~Aldam, T.~Roch, E.A. Brener, E.~Bouchbinder, and J.-F. Molinari.
\newblock The emergence of crack-like behavior of frictional rupture: Edge
  singularity and energy balance.
\newblock {\em Earth Planet. Sci. Lett.}, 531:115978, 2020.

\bibitem{SvetFineberg}
I.~Svetlizky and J.~Fineberg.
\newblock Classical shear cracks drive the onset of dry frictional motion.
\newblock {\em Nature}, 509:205, 2014.

\bibitem{Norw3}
K.~Th\o{}gersen, J.~Tr\o{}mborg, H.A. Sveinsson, A.~Malthe-S\o{}renssen, and
  J.~Scheibert.
\newblock History-dependent friction and slow slip from time-dependent
  microscopic junction laws studied in a statistical framework.
\newblock {\em Phys. Rev. E}, 89:052401, 2014.

\bibitem{Molinari_frac}
J.~Kammer, M.~Radiguet, J.~Ampuero, and J.F. Molinari.
\newblock Linear elastic fracture mechanics predicts the propagation distance
  of frictional slip.
\newblock {\em Tribol. Lett.}, 57:23, 2015.

\bibitem{Fineberg_frac}
E.~Bayart, I.~Svetlizky, and J.~Fineberg.
\newblock Fracture mechanics determine the lengths of interface ruptures that
  mediate frictional motion.
\newblock {\em Nature Physics}, 12:166, 2016.

\bibitem{GvirFineberg}
S.~Gvirtzman and J.~Fineberg.
\newblock Nucleation fronts ignite the interface rupture that initiates
  frictional motion.
\newblock {\em Nature Phys.}, 17:1037, 2021.

\bibitem{LabonteInsects}
D.~Labonte, J.A. Williams, and W.~Federle.
\newblock Surface contact and design of fibrillar 'friction pads' in stick
  insects (carausius morosus): mechanisms for large friction coefficients and
  negligible adhesion.
\newblock {\em J. R. Soc. Interface}, 11:0034, 2014.

\bibitem{GorbSnake}
A.~Filippov and Gorb S.N.
\newblock Modelling of the frictional behaviour of the snake skin covered by
  anisotropic surface nanostructures.
\newblock {\em Sci. Rep.}, 6:23539, 2016.

\bibitem{Tramsen}
H.T. Tramsen, S.N. Gorb, H.~Zhang, P.~Manoonpong, Z.~Dai, and L.~Heepe.
\newblock Inversion of friction anisotropy in a bio-inspired asymmetrically
  structured surface.
\newblock {\em J. Royal Soc. Interface}, 15:2017062, 2018.

\bibitem{Etsion}
I.~Etsion.
\newblock Improving tribological performance of mechanical components by laser
  surface texturing.
\newblock {\em Tribol. Lett.}, 17:733, 2004.

\bibitem{Gachot}
C.~Gachot, Rosenkranz A., Hsu S.M., and H.L. Costa.
\newblock A critical assessment of surface texturing for friction and wear
  improvement.
\newblock {\em Wear}, 372:21, 2017.

\bibitem{Rosenkranz}
A.~Rosenkranz, P.G. Grutzmacher, C.~Gachot, and H.L. Costa.
\newblock Surface texturing in machine elements - a critical discussion for
  rolling and sliding contacts.
\newblock {\em Adv. Eng. Mater.}, 21:1900194, 2019.

\bibitem{Dieterich}
J.H. Dieterich.
\newblock Preseismic fault slip and earthquake prediction.
\newblock {\em J. Geophys. Res.}, 83:3940, 1978.

\bibitem{Rice93}
J.R. Rice.
\newblock Spatio‐temporal complexity of slip on a fault.
\newblock {\em J. Geophys. Res.}, 98:9885, 1993.

\bibitem{LapustaRice}
N.~Lapusta and J.R. Rice.
\newblock Nucleation and early seismic propagation of small and large events in
  a crustal earthquake model.
\newblock {\em J. Geophys. Res.}, 108:2205, 2003.

\bibitem{Bouchbinder_geo}
Y.~Bar~Sinai, E.A. Brener, and E.~Bouchbinder.
\newblock Slow rupture of frictional interfaces.
\newblock {\em Geophys. Res. Lett.}, 39:L03308, 2012.

\bibitem{Lambert}
L.~Lambert, N.~Lapusta, and Perry S.
\newblock Propagation of large earthquakes as self-healing pulses or mild
  cracks.
\newblock {\em Nature}, 591:252, 2021.

\bibitem{Our2D}
G.~Costagliola, F.~Bosia, and N.M. Pugno.
\newblock A 2-d model for friction of complex anisotropic surfaces.
\newblock {\em J. Mech. Phy. Solids}, 112:50, 2018.

\bibitem{Prodanov}
N.~Prodanov, C.~Gachot, A.~Rosenkranz, F.~Muklich, and M.~Muser.
\newblock Contact mechanics of laser-textured surfaces.
\newblock {\em Tribol. Lett.}, 50:41, 2013.

\bibitem{Gachot_cit}
C.~Gachot, A.~Rosenkranz, L.~Reinert, E.~Ramos-Moore, N.~Souza, M.~Muser, and
  F.~Muklich.
\newblock Dry friction between laser-patterned surfaces: Role of alignment,
  structural wavelength and surface chemistry.
\newblock {\em Tribol. Lett.}, 49:193, 2013.

\bibitem{Rosenkranz_cit}
A.~Rosenkranz, L.~Reinert, C.~Gachot, and F.~Mücklich.
\newblock Alignment and wear debris effects between laser-patterned steel
  surfaces under dry sliding conditions.
\newblock {\em Wear}, 318:49, 2014.

\bibitem{Valeri}
I.~Gnilitskyi, A.~Rota, E.~Gualtieri, S.~Valeri, and L.~Orazi.
\newblock Tribological properties of high-speed uniform femtosecond laser
  patterning on stainless steel.
\newblock {\em Lubricants}, 7:83, 2019.

\bibitem{Absi}
E.~Absi and W.~Prager.
\newblock Comparison of equivalence and finite element methods.
\newblock {\em Comp. Methods. in Appl. Mech. Eng.}, 6:59, 1975.

\bibitem{Our1D}
G.~Costagliola, F.~Bosia, and N.M. Pugno.
\newblock Static and dynamic friction of hierarchical surfaces.
\newblock {\em Phys. Rev. E}, 94:063003, 2016.

\bibitem{Balestra}
S.~Balestra, G.~Costagliola, A.~Pegoraro, F.~Picollo, J.-F. Molinari, N.M.
  Pugno, E.~Vittone, F.~Bosia, and A.~Sin.
\newblock Experimental and numerical study of the effect of surface patterning
  on the frictional properties of polymer surfaces.
\newblock {\em J. Tribol.}, 144:031704, 2022.

\bibitem{Berardo}
A.~Berardo, G.~Costagliola, S.~Ghio, M.~Boscardin, F.~Bosia, and N.M. Pugno.
\newblock An experimental-numerical study of the adhesive static and dynamic
  friction of micropatterned soft polymer surfaces.
\newblock {\em Materials \& Design}, 181:107930, 2019.

\bibitem{Scheibert}
R.~Sahli, G.~Pallares, C.~Ducottet, I.~E. Ben~Ali, S.~Al~Akhrass, M.~Guibert,
  and J.~Scheibert.
\newblock Evolution of real contact area under shear and the value of static
  friction of soft materials.
\newblock {\em Proc. Natl. Acad. Sci.}, 115:471, 2018.

\bibitem{Maegawa_patt}
S.~Maegawa, F.~Itoigawa, and T.~Nakamura.
\newblock Effect of surface grooves on kinetic friction of a rubber slider.
\newblock {\em Tribol. Int.}, 102:326, 2016.

\bibitem{Guarino}
R.~Guarino, G.~Costagliola, F.~Bosia, and N.M. Pugno.
\newblock Evidence of friction reduction in laterally graded materials.
\newblock {\em Beil. J. Nanotechnol.}, 9:2443, 2018.

\bibitem{Varenberg}
B.~Murarash, Y.~Itovicha, and M.~Varenberg.
\newblock Tuning elastomer friction by hexagonal surface patterning.
\newblock {\em Soft Matters}, 7:5553, 2011.

\bibitem{Li}
N.~Li, E.~Xu, Z.~Liu, X.~Wang, and L.~Liu.
\newblock Tuning apparent friction coefficient by controlled patterning bulk
  metallic glasses surfaces.
\newblock {\em Sci. Rep.}, 6:39388, 2016.

\bibitem{Lucas}
L.~Brely, F.~Bosia, and N.M. Pugno.
\newblock Emergence of the interplay between hierarchy and contact splitting in
  biological adhesion highlighted through a hierarchical shear lag model.
\newblock {\em Soft Matter}, 14:5509, 2018.

\bibitem{Gecko_Gao}
B.~Chen, P.~Wu, and H.~Gao.
\newblock Pre-tension generates strongly reversible adhesion of a spatula pad
  on substrate.
\newblock {\em J. R. Soc. Interface}, 6:529, 2009.

\bibitem{Papadimitri}
P.~Papadimitriou.
\newblock Identification of seismic precursors before large
  earthquakes:decelerating and accelerating seismic patterns.
\newblock {\em J. Geophys. Res.}, 113:B04306, 2008.

\bibitem{Sykes}
L.R. Sykes.
\newblock Intermediate- and long-term earthquake prediction.
\newblock {\em Proc. Natl. Acad. Sci.}, 93:3732, 1996.

\bibitem{Rundle}
J.B. Rundle, J.R. Holliday, M.~Yoder, M.K. Sachs, A.~Donnellan, D.L. Turcotte,
  K.F. Tiampo, W.~Klein, and L.H. Kellogg.
\newblock Earthquake precursors: activation or quiescence?
\newblock {\em Geophys. J. Int.}, 187:225, 2011.

\bibitem{Linde_Aseismic}
A.T. Linde, K.~Suyehiro, S.~Miura, I.S. Sacks, and A.~Takagi.
\newblock Episodic aseismic earthquake precursors.
\newblock {\em Nature}, 334:513, 1988.

\bibitem{Schmittbuhl_Izmit}
M.~Bouchon, H.~Karabulut, M.~Aktar, S.~Ozalaybey, J.~Schmittbuhl, and M.~Bouin.
\newblock Extended nucleation of the 1999 mw 7.6 izmit earthquake.
\newblock {\em Science}, 331:877, 2011.

\bibitem{Schmittbuhl_Precurs}
M.~Bouchon, V.~Durand, D.~Marsan, H.~Karabulut, and Schmittbuhl J.
\newblock The long precursory phase of most large interplate earthquakes.
\newblock {\em Nat. Geosci.}, 6:299, 2013.

\bibitem{Hasegawa}
A.~Hasegawa and K.~Yoshida.
\newblock Preceding seismic activity and slow slip events in the source area of
  the 2011 mw 9.0 tohoku-oki earthquake: a review.
\newblock {\em Geosci. Lett.}, 2:6, 2013.

\bibitem{Ellsworth_1}
W.L. Ellsworth and G.C. Beroza.
\newblock Seismic evidence for an earthquake nucleation phase.
\newblock {\em Science}, 268:851, 1995.

\bibitem{Ellsworth_2}
D.A. Dodge, G.C. Beroza, and W.~L. Ellsworth.
\newblock Detailed observations of california foreshock sequences: implications
  for the earthquake initiation process.
\newblock {\em J. Geophys. Res.}, 101:371, 1996.

\bibitem{Ellsworth_3}
D.P. Schaff, G.H.R. Bokelmann, G.C. Beroza, F.~Waldhauser, and W.L. Ellsworth.
\newblock High-resolution image of calaveras fault seismicity.
\newblock {\em J. Geophys. Res.}, 107:2186, 2002.

\bibitem{Lapusta}
N.~Lapusta, J.R. Rice, Y.~Ben-Zion, and G.~Zheng.
\newblock Elastodynamic analysis for slow tectonic loading with spontaneous
  rupture episodes on faults with rate- and state-dependent friction.
\newblock {\em J. Geophys. Res.}, 105:765, 2000.

\bibitem{Ferdowsi}
B.~Ferdowsi, M.~Griffa, R.A. Guyer, P.A. Johnson, C.~Marone, and J.~Carmeliet.
\newblock Microslips as precursors of large slip events in the stick-slip
  dynamics of sheared granular layers: A discrete element model analysis.
\newblock {\em Geophys. Res. Lett.}, 40:4194, 2013.

\bibitem{Brown_SB2D}
S.R. Brown, C.H. Scholz, and J.B. Rundle.
\newblock A simplified spring-block model of earthquakes.
\newblock {\em Geophys. Res. Lett.}, 18:215, 1991.

\bibitem{Mori_1}
T.~Mori and H.~Kawamura.
\newblock Simulation study of the two-dimensional burridge-knopoff model of
  earthquakes.
\newblock {\em J. Geophys. Res.}, 113:B06301, 2008.

\bibitem{Mori_2}
T.~Mori and H.~Kawamura.
\newblock Simulation study of earthquakes based on the two-dimensional
  burridge-knopoff model with long-range interactions.
\newblock {\em Phys. Rev. E}, 77:051123, 2008.

\bibitem{Ohnaka}
N.~Ohnaka and L.~Shen.
\newblock Scaling of the shear rupture process from nucleation to dynamic
  propagation: Implications of geometric irregularity of the rupturing
  surfaces.
\newblock {\em J. Geophys. Res.}, 104:817, 1999.

\bibitem{Aochi}
H.~Aochi and S.~Ide.
\newblock Role of multiscale heterogeneity in fault slip from quasi-static
  numerical simulations.
\newblock {\em Earth Planets Space}, 69:94, 2017.

\bibitem{Stein}
R.S. Stein, G.C.P. King, and J.~Lin.
\newblock Stress triggering of the 1994 m = 6.7 northridge, california,
  earthquake by its predecessor.
\newblock {\em Science}, 265:1432, 1994.

\end{thebibliography}

\end{document}